\documentclass{aa}  

\usepackage{graphicx}
\usepackage{amsmath,amsfonts,amssymb}
\usepackage{natbib}

\usepackage{txfonts}
\usepackage{xcolor}

\usepackage{blindtext}
\usepackage{float}
\usepackage{dblfloatfix}
\usepackage{afterpage}
\usepackage{ifthen}
\usepackage[morefloats=12]{morefloats}

\usepackage{placeins}
\usepackage{multicol}
\usepackage[export]{adjustbox}\usepackage[breaklinks,colorlinks,citecolor=blue]{hyperref}
\bibpunct{(}{)}{;}{a}{}{,}
\usepackage[switch]{lineno}
\definecolor{linkcolor}{rgb}{0.6,0,0}
\definecolor{citecolor}{rgb}{0,0,0.75}
\definecolor{urlcolor}{rgb}{0.12,0.46,0.7}
\hypersetup{linktocpage}
\usepackage{bold-extra}
\usepackage{tabularx, booktabs}

\def\setsymbol#1#2{\expandafter\def\csname #1\endcsname{#2}}
\def\getsymbol#1{\csname #1\endcsname}

\def\Planck{\textit{Planck}}

\newbox\tablebox    \newdimen\tablewidth
\def\leaderfil{\leaders\hbox to 5pt{\hss.\hss}\hfil}
\def\tablenote#1 #2\par{\begingroup \parindent=0.8em
    \abovedisplayshortskip=0pt\belowdisplayshortskip=0pt
    \noindent
    $$\hss\vbox{\hsize\tablewidth \hangindent=\parindent \hangafter=1 \noindent
    \hbox to \parindent{$^#1$\hss}\strut#2\strut\par}\hss$$
    \endgroup}

\def\L2{\ifmmode L_2\else $L_2$\fi}

\def\DeltaT{\ifmmode \Delta T\else $\Delta T$\fi}
\def\deltat{\ifmmode \Delta t\else $\Delta t$\fi}
\def\fknee{\ifmmode f_{\rm knee}\else $f_{\rm knee}$\fi}
\def\Fmax{\ifmmode F_{\rm max}\else $F_{\rm max}$\fi}
\def\solar{\ifmmode{\rm M}_{\mathord\odot}\else${\rm M}_{\mathord\odot}$\fi}
\def\Msolar{\ifmmode{\rm M}_{\mathord\odot}\else${\rm M}_{\mathord\odot}$\fi}
\def\Lsolar{\ifmmode{\rm L}_{\mathord\odot}\else${\rm L}_{\mathord\odot}$\fi}
\def\inv{\ifmmode^{-1}\else$^{-1}$\fi}
\def\mo{\ifmmode^{-1}\else$^{-1}$\fi}
\def\sup#1{\ifmmode ^{\rm #1}\else $^{\rm #1}$\fi}
\def\expo#1{\ifmmode \times 10^{#1}\else $\times 10^{#1}$\fi}
\def\,{\thinspace}
\def\lsim{\mathrel{\raise .4ex\hbox{\rlap{$<$}\lower 1.2ex\hbox{$\sim$}}}}
\def\gsim{\mathrel{\raise .4ex\hbox{\rlap{$>$}\lower 1.2ex\hbox{$\sim$}}}}

\def\simprop{\mathrel{\raise .4ex\hbox{\rlap{$\propto$}\lower 1.2ex\hbox{$\sim$}}}}
\def\deg{\ifmmode^\circ\else$^\circ$\fi}
\def\pdeg{\ifmmode $\setbox0=\hbox{$^{\circ}$}\rlap{\hskip.11\wd0 .}$^{\circ}
          \else \setbox0=\hbox{$^{\circ}$}\rlap{\hskip.11\wd0 .}$^{\circ}$\fi}
\def\arcs{\ifmmode {^{\scriptstyle\prime\prime}}
          \else $^{\scriptstyle\prime\prime}$\fi}
\def\arcm{\ifmmode {^{\scriptstyle\prime}}
          \else $^{\scriptstyle\prime}$\fi}
\newdimen\sa  \newdimen\sb
\def\parcs{\sa=.07em \sb=.03em
     \ifmmode \hbox{\rlap{.}}^{\scriptstyle\prime\kern -\sb\prime}\hbox{\kern -\sa}
     \else \rlap{.}$^{\scriptstyle\prime\kern -\sb\prime}$\kern -\sa\fi}
\def\parcm{\sa=.08em \sb=.03em
     \ifmmode \hbox{\rlap{.}\kern\sa}^{\scriptstyle\prime}\hbox{\kern-\sb}
     \else \rlap{.}\kern\sa$^{\scriptstyle\prime}$\kern-\sb\fi}
\def\ra[#1 #2 #3.#4]{#1\sup{h}#2\sup{m}#3\sup{s}\llap.#4}
\def\dec[#1 #2 #3.#4]{#1\deg#2\arcm#3\arcs\llap.#4}
\def\deco[#1 #2 #3]{#1\deg#2\arcm#3\arcs}
\def\rra[#1 #2]{#1\sup{h}#2\sup{m}}

\def\dots{\relax\ifmmode \ldots\else $\ldots$\fi}
\def\WHzsr{\ifmmode $W\,Hz\mo\,sr\mo$\else W\,Hz\mo\,sr\mo\fi}
\def\mHz{\ifmmode $\,mHz$\else \,mHz\fi}
\def\GHz{\ifmmode $\,GHz$\else \,GHz\fi}
\def\mKs{\ifmmode $\,mK\,s$^{1/2}\else \,mK\,s$^{1/2}$\fi}
\def\muKs{\ifmmode \,\mu$K\,s$^{1/2}\else \,$\mu$K\,s$^{1/2}$\fi}
\def\muKRJs{\ifmmode \,\mu$K$_{\rm RJ}$\,s$^{1/2}\else \,$\mu$K$_{\rm RJ}$\,s$^{1/2}$\fi}
\def\muKHz{\ifmmode \,\mu$K\,Hz$^{-1/2}\else \,$\mu$K\,Hz$^{-1/2}$\fi}
\def\MJysr{\ifmmode \,$MJy\,sr\mo$\else \,MJy\,sr\mo\fi}
\def\MJysrmK{\ifmmode \,$MJy\,sr\mo$\,mK$_{\rm CMB}\mo\else \,MJy\,sr\mo\,mK$_{\rm CMB}\mo$\fi}
\def\microns{\ifmmode \,\mu$m$\else \,$\mu$m\fi}

\def\muK{\ifmmode \,\mu$K$\else \,$\mu$\hbox{K}\fi}
\def\microK{\ifmmode \,\mu$K$\else \,$\mu$\hbox{K}\fi}
\def\muW{\ifmmode \,\mu$W$\else \,$\mu$\hbox{W}\fi}
\def\kms{\ifmmode $\,km\,s$^{-1}\else \,km\,s$^{-1}$\fi}
\def\kmsMpc{\ifmmode $\,\kms\,Mpc\mo$\else \,\kms\,Mpc\mo\fi}
\providecommand{\sorthelp}[1]{}

\def\Cosmoglobe{\textsc{Cosmoglobe}}
\def\Planck{\textit{Planck}}

\def\COBE{\textit{COBE}}
\def\GAIA{\textit{Gaia}}
\def\gaia{\textit{Gaia}}
\def\Gaia{\textit{Gaia}}
\def\WISE{WISE}

\def\IRAS{\textit{{IRAS}}}

\newcommand{\dv}[0]{\vec{d}}
\renewcommand{\t}[0]{\vec{t}}
\newcommand{\A}[0]{\tens{A}}
\newcommand{\B}[0]{\tens{B}}

\newcommand{\G}[0]{\tens{G}}
\newcommand{\n}[0]{\vec{n}}

\newcommand{\s}[0]{\vec{s}}
\renewcommand{\a}[0]{\vec{a}}

\renewcommand{\L}[0]{\tens{L}}

\newcommand{\N}[0]{\tens{N}}
\newcommand{\M}[0]{\tens{M}}

\renewcommand{\r}[0]{\vec{r}}

\renewcommand{\P}[0]{\tens{P}}

\usepackage{lineno}
\begin{document}

   \title{\bfseries{\Cosmoglobe\ DR2. IV. Modelling starlight\\ in DIRBE with \GAIA\ and \WISE}}

   \newcommand{\oslo}[0]{1}
\newcommand{\milan}[0]{2}
\newcommand{\ijclab}[0]{3}
\newcommand{\gothenberg}[0]{4}
\newcommand{\trento}[0]{5}
\newcommand{\milanoinfn}[0]{6}
\author{\small
M.~Galloway\inst{\oslo}\thanks{Corresponding author: M.~Galloway; \url{mathew.galloway@astro.uio.no}}
\and
E.~Gjerl\o w\inst{\oslo}
\and
M.~San\inst{\oslo}
\and
R.~M.~Sullivan\inst{\oslo}
\and
D.~J.~Watts\inst{\oslo}
\and
R.~Aurvik\inst{\oslo}
\and
A.~Basyrov\inst{\oslo}
\and
L.~A.~Bianchi\inst{\oslo}
\and
A.~Bonato\inst{\milan}
\and
M.~Brilenkov\inst{\oslo}
\and
H.~K.~Eriksen\inst{\oslo}
\and
U.~Fuskeland\inst{\oslo}
\and
K.~A.~Glasscock\inst{\oslo}
\and
L.~T.~Hergt\inst{\ijclab}
\and
D.~Herman\inst{\oslo}
\and
J.~G.~S.~Lunde\inst{\oslo}
\and
A.~I.~Silva Martins\inst{\oslo}
\and
D.~Sponseller\inst{\gothenberg}
\and
N.-O.~Stutzer\inst{\oslo}
\and
H.~Thommesen\inst{\oslo}
\and
V.~Vikenes\inst{\oslo}
\and
I.~K.~Wehus\inst{\oslo}
\and
L.~Zapelli\inst{\milan, \trento, \milanoinfn}
}
\institute{\small
Institute of Theoretical Astrophysics, University of Oslo, Blindern, Oslo, Norway\goodbreak
\and
Dipartimento di Fisica, Università degli Studi di Milano, Via Celoria, 16, Milano, Italy
\and
Laboratoire de Physique des 2 infinis -- Irène Joliot Curie (IJCLab), Orsay, France
\and
Department of Space, Earth and Environment, Chalmers University of Technology, Gothenburg, Sweden\goodbreak
\and
Università di Trento, Università degli Studi di Milano, CUP E66E23000110001\goodbreak
\and
INFN sezione di Milano, 20133 Milano, Italy\goodbreak
}

   %\institute{Institute of Theoretical Astrophysics, University of Oslo, Blindern, Oslo, Norway}
  
   % Shortened title, author list for top of page 
   \titlerunning{Compact objects in DIRBE}
   \authorrunning{Galloway et al.}

   \date{\today} 
   
   \abstract{We present a model of starlight emission in the Diffuse Infrared Background Explorer (DIRBE) data between 1.25 and 25$\,\mu$m based on \textit{Gaia} and WISE measurements. We include two classes of compact objects, namely bright stars with individual spectral energy densities (SEDs) measured by \textit{Gaia}, and a combined diffuse background of dim point source emission. Of the 424\ 829 bright sources that we fit, the number of stars with a flux density detected by WISE at Galactic latitudes $|b|>20^{\circ}$ at more than $5\,\sigma$ is 94\,680, for an average of 1.36~stars per DIRBE beam area. For each star, we adopt physical parameters ($T_{\mathrm{eff}}$, $\log g$, and [M/H]) from \textit{Gaia}; use these to identify a best-fit effective SED with the PHOENIX stellar model library; convolve with the respective DIRBE bandpass; and fit an overall free amplitude per star within the Bayesian end-to-end \Cosmoglobe\ DR2 framework. The contributions from faint sources are accounted for by coadding all 710\ 825\ 587 WISE sources not included as bright stars, and fit one single overall amplitude per DIRBE band. Based on this model we find that total star emission accounts for 91\,\% of the observed flux density at 2.2\,$\mu$m; 54\,\% at 4.9$\,\mu$m; and 1\,\% at 25\,$\mu$m. As shown in companion papers, this new model is sufficiently accurate to support high-precision measurements of both the Cosmic Infrared Background monopole and zodiacal light emission in the three highest DIRBE frequencies.}


   \maketitle

\setcounter{tocdepth}{2}
   
% INTRODUCTION
%-------------------------------------------------------------------
\section{Introduction}
%\the\textwidth \the\columnwidth

Modelling the astrophysical sky using the \COBE-Diffuse Infrared Background Explorer (DIRBE; \citealp{DIRBE}) data from 1 to 100\,000 GHz requires an understanding of all the various components that make it up. At the highest of these frequencies, the largest power contribution comes from resolved stars in our galaxy, as well as an unresolved background of dimmer sources. The four shortest wavelength/highest frequency DIRBE bands between 1.25 and 4.9\,$\mu$m are all dominated by star emission \citep{arendt1998}, making accurate star modelling essential for using these data in combination with others in a comprehensive Bayesian model of the large-scale infrared sky \citep{CG02_01}. Additionally, point sources are also subdominant contributors in the DIRBE 12 and 25$\mu$m channels, which if not handled correctly could bias derived constraints on the zodiacal light and other components \citep{K98,CG02_02}.

Many full-sky datasets exist which measure the emissions from point sources in our galaxy and beyond. Starting with \IRAS\ in 1983 \citep{neugebauer:1984}, there have been several ground and space missions to map the infrared sky and put constraints on star formation and evolution, the Cosmic Infrared Background (CIB) and Active Galactic Nuclei (AGN), including the Infrared Space Observatory (ISO; \citealp{iso}), Spitzer \citep{spitzer}, the 2-Micron All-Sky Survey (2MASS) \citep{2mass}, and the current-generation James Webb Space Telescope (JWST) \citep{jwst}. For the purposes of compatibility with the DIRBE data, however, the most useful external datasets are the Wide-Field Infrared Survey Explorer (WISE) \citep{wise}, and \GAIA, which produced high precision star observations at optical wavelengths \citep{gaia, gaia2}. 

In this paper, which is part of the larger \Cosmoglobe\ DR2 results paper suite \citep[][and references therein]{CG02_01}, we discuss the modelling of compact objects in DIRBE as part of our larger full sky astrophysical model. Section \ref{sec:models} discusses this modelling in a larger Bayesian context. Section \ref{sec:results} shows the astrophysical results of the works by presenting a unified star model and its associated errors. Section \ref{sec:consistency} then shows the consistency between our results and other work in this area. Finally, we conclude in Sect.~\ref{sec:conclusions}, discussing the impacts of this work and offering some avenues for future research.

\section{Bayesian star modelling in \Cosmoglobe\ DR2}
\label{sec:models}

\begin{figure}
  \centering
  \includegraphics[width=\columnwidth]{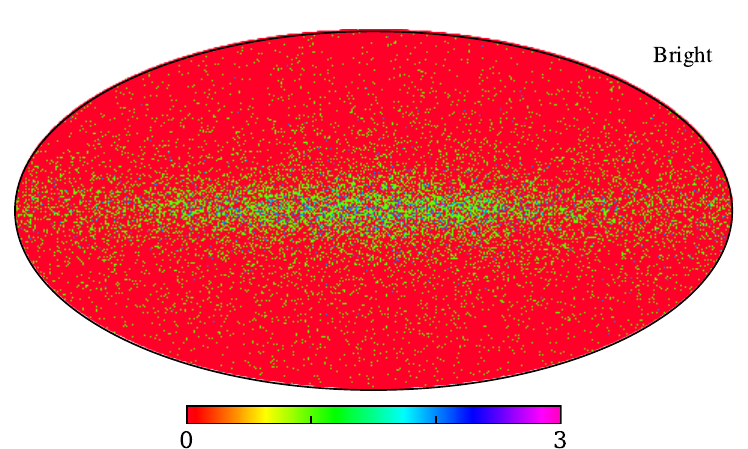}\\
  \includegraphics[width=\columnwidth]{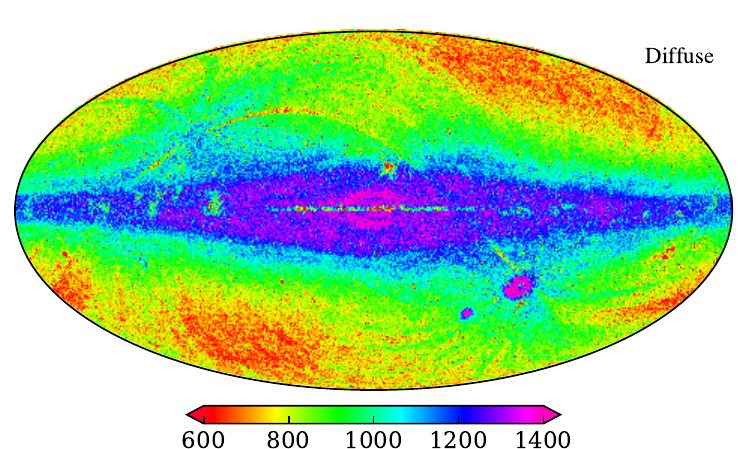}
  \caption{Top: Total number of bright sources in each pixel. Bottom: Total number of sources included in the diffuse template, which shows a clear imprint of the WISE scan strategy.}
  \label{fig:starcount}
\end{figure}

\begin{figure}
  \centering
  \includegraphics[width=\columnwidth]{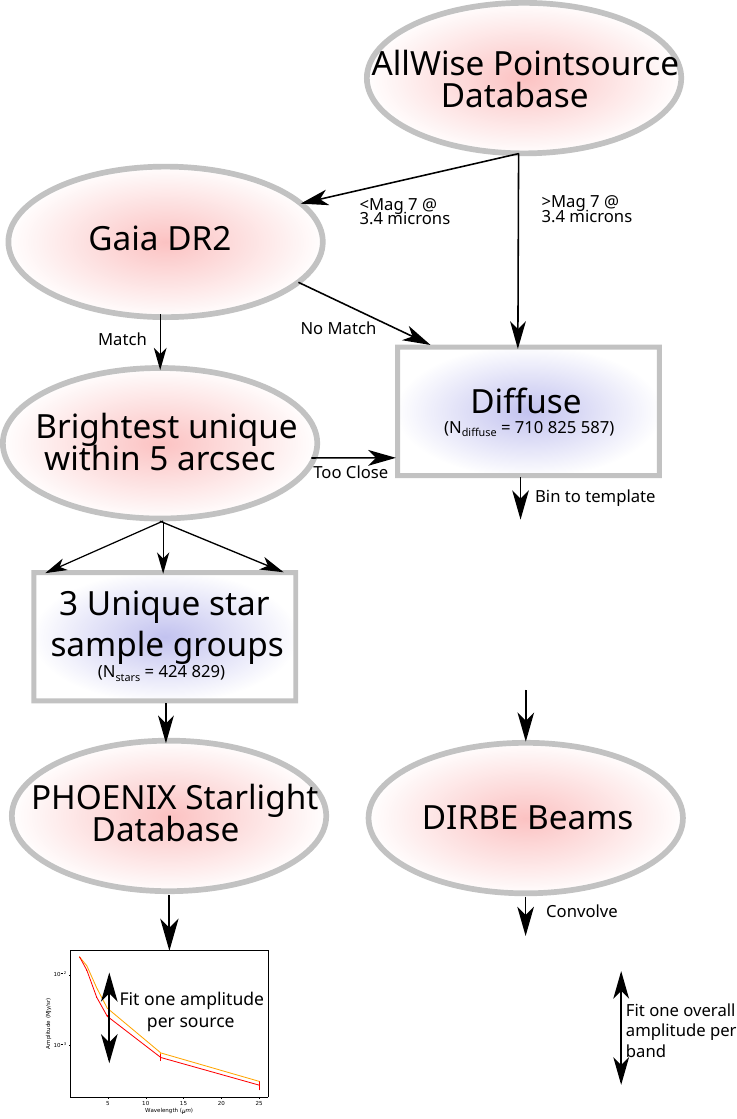}\\
  \caption{Schematic diagram of the \Cosmoglobe\ DR2 compact object processing pipeline. Red ovals represent input data, and the blue boxes indicate the two classes of sources described in this paper. }
  \label{fig:diagram}
\end{figure}

In this work, we define two categories of stars, which are treated differently but jointly. The first category includes objects in the AllWISE 3.4 $\,\mu$m catalog \citep{wiseCat} with a magnitude brighter than 7, but excluding sources that are spatially too close to distinguish with the DIRBE beam. This comprises a total of 424\,829 bright objects, and the top panel of Fig.~\ref{fig:starcount} shows the distribution of these across the sky in terms of the total number of stars per HEALPix\footnote{\url{http://healpix.sourceforge.net}} map \citep{healpix} $7'$ pixel ($N_\mathrm{side}=512$). Each of these stars are modelled individually using data from \Gaia\ in the current analysis, as described in Sect.~\ref{sec:starmodel}. The second category includes all the other sources in the AllWISE catalog, which comprises a total of 710\,825\,587 objects. The distribution of these are shown in the bottom panel of Fig.~\ref{fig:starcount}. In the current analysis, these are all co-added into one effective spatial template of the dimmer ``background'', and we will hereafter refer to this as the ``diffuse'' component; for details, see Sect.~\ref{sec:diffusemodel}. An overview of the full compact object processing pipeline used in this analysis is shown in Fig.~\ref{fig:diagram}. 

\subsection{Data model and sampling algorithm}
\label{sec:model}

\begin{figure}
  \includegraphics[width=0.96\columnwidth]{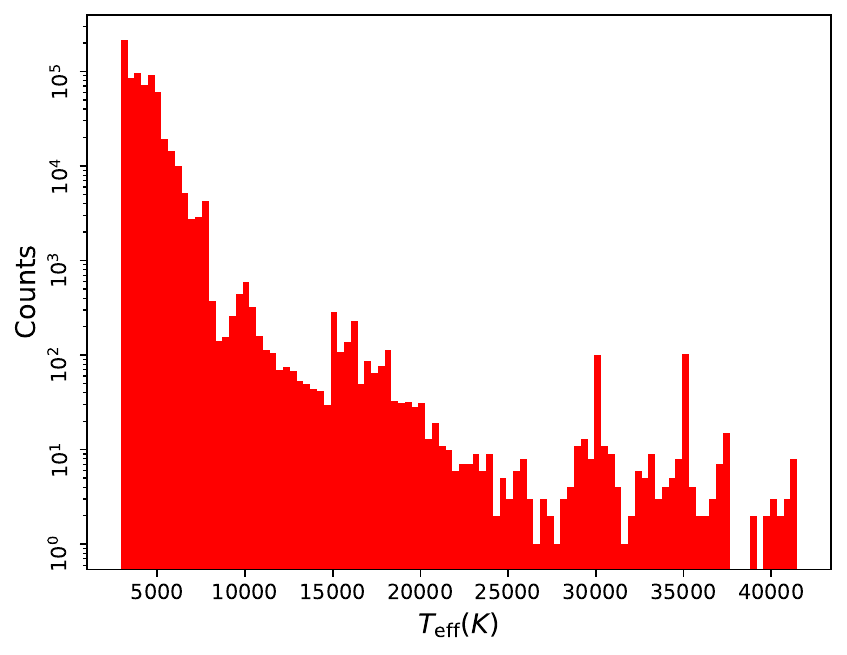}
  \includegraphics[width=0.96\columnwidth]{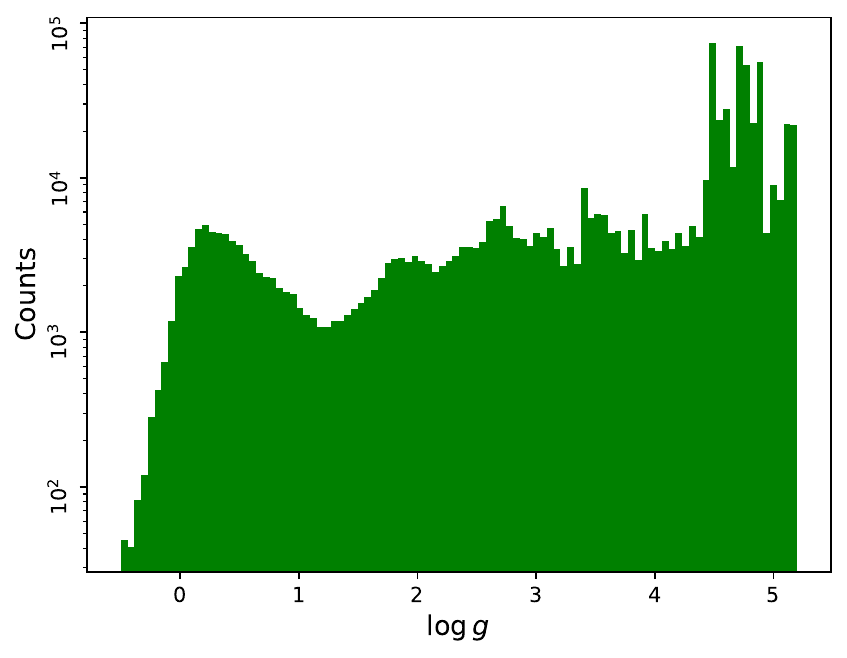}\\
  \includegraphics[width=0.96\columnwidth]{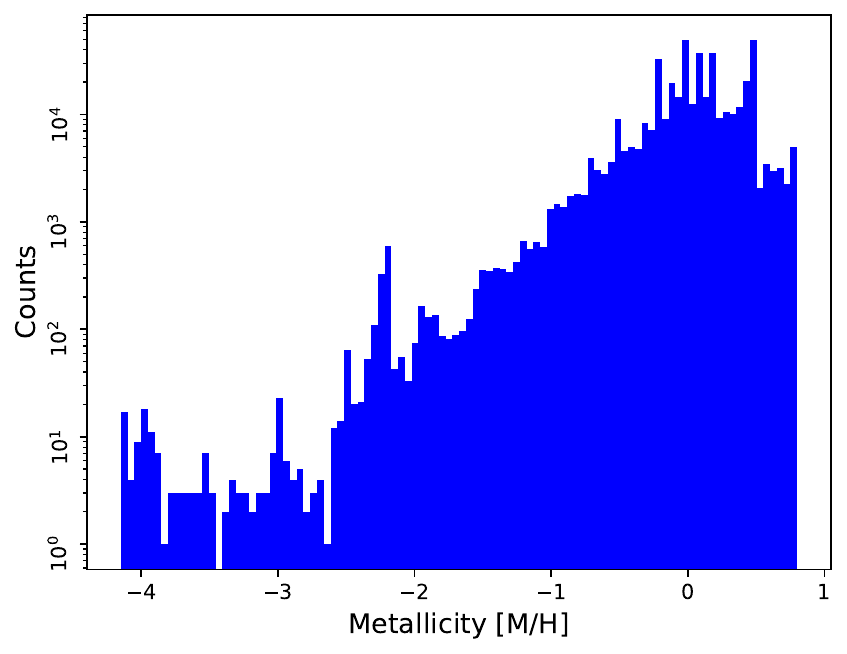}\\
  \caption{Histograms of the three star parameters taken from \Gaia\ (from top to bottom: effective temperature, gravitational acceleration, and metallicity) that are used by the PHOENIX database to determine the star SEDs, for the 424\,829 bright stars used in this analysis.}
  \label{fig:gaiacat}
\end{figure}

The objective of this paper is to fit the amplitude of both individual bright stars and the overall diffuse background jointly as part of the global Bayesian \Cosmoglobe\ DR2 analysis framework. The details of this process are discussed by \citet{CG02_01, CG02_02, CG02_07}, and here we only give a brief review of the main points that are directly relevant for starlight modelling.

The basic \Cosmoglobe\ DR2 data model takes the following form,
\begin{align}
	\label{eq:model}
	\dv &=\G\P\B\sum_{c=1}^{n_{\mathrm{comp}}}\M_c\a_c+\s_{\mathrm{zodi}} +
          \s_{\mathrm{static}} + \n_\mathrm{corr} + \n_\mathrm w\\
        &\equiv \s_{\mathrm{tot}} + \n_{\mathrm{w}},
\end{align}
where $\dv$ indicates time-ordered data; $\G$ represents instrumental gain; $\P$ encodes the telescope pointing; $\B$ denotes beam convolution; $\M_c$ is a so-called mixing matrix that defines the relative strength of component $c$ at frequency $\nu$; $\a_c$ is a reference component amplitude; $\s_{\mathrm{zodi}}$ and $\s_{\mathrm{static}}$ are contributions from zodiacal light; and $\n_{\mathrm{corr}}$ and $\n_{\mathrm{w}}$ are correlated and white instrumental noise, respectively. We note that not all parameters are fitted for all channels; see \citet{CG02_01} for details.

As far as this paper is concerned, the key element is $\M_c\a_c$, which defines the astrophysical sky model. In general, this expression includes terms for each relevant physical process, which for the current analysis are thermal dust \citep{CG02_05,CG02_06,CG02_07}, free-free \citep{dickinson2003,planck2014-a12}, CIB (monopoles; \citealp{DIRBE,CG02_03}), and the two starlight populations introduced above. Explicitly, we adopt the following data model for the starlight emission in the following,
\begin{align}
  \M_{\mathrm{stars}}\a_{\mathrm{stars}} &= \sum_{j=1}^{n_{\mathrm{s}}} \M^i_{\mathrm{bright}}\a^i_{\mathrm{bright}} + \a_{\mathrm{diff}}\t_{\mathrm{diff}}\\
  &= U^i_{\mathrm{mJy}} \sum_{j=1}^{n_{\mathrm{s}}} \epsilon_j(\nu)\,f_{\mathit{Gaia},j} \a_{\mathrm{s},j} + \a_{\mathrm{diff}}\t_{\mathrm{diff}},
  \label{eq:datamodel}
\end{align}
where $U_{\mathrm{mJy}}$ is a unit conversion factor;
$\epsilon_j(\nu)$ is a frequency-dependent extinction factor;
$f_{\mathit{Gaia}}$ is a (bandpass-convolved) Spectral Energy Density (SED) amplitude per
channel and source; and $\a_{\mathrm{s},j}$ is the amplitude per
source, which is the only parameter that is fitted freely per
object. Finally, the second term simply consists of a single overall
diffuse template, $\t_{\mathrm{diff}}$, multiplied with a free
template amplitude per channel. The definition and construction of
each of these factors are described in Sects.~\ref{sec:starmodel} and
\ref{sec:diffusemodel}.

As noted above, the \Cosmoglobe\ DR2 algorithm is a Gibbs sampler, which in practice means that we sample one parameter in Eq.~\ref{eq:model} at a time, while keeping all others temporarily fixed, and then iterate through all parameters. As far as the current paper is concerned, the key steps are therefore to be able to sample from the two following conditional distributions,
\begin{align}
  \a_{\mathrm{bright}} &\leftarrow P(\a_{\mathrm{bright}}| \dv, \G, \M_{\mathrm{dust}}, \a_{\mathrm{dust}}, \s_{\mathrm{zodi}}, \s_{\mathrm{static}}, \n_{\mathrm{corr}}, a_{\mathrm{diff}}), \\
  \a_{\mathrm{diff}} &\leftarrow P(\a_{\mathrm{diff}}|  \dv, \G, \M_{\mathrm{dust}}, \a_{\mathrm{dust}}, \s_{\mathrm{zodi}}, \s_{\mathrm{static}}, \n_{\mathrm{corr}}, a_{\mathrm{bright}}).
\end{align}
As described by \citet{CG02_01}, in practice this is achieved by 1) subtracting the time-domain correction terms ($\s_{\mathrm{zodi}}$, $\s_{\mathrm{static}}$, and $\n_{\mathrm{corr}}$) from the raw time-ordered data; 2) binning the resulting cleaned data into pixelized maps; 3) subtracting all other astrophysical components except the starlight; and 4) drawing $\a_{\mathrm{bright}}$ and $\a_{\mathrm{diff}}$ from their  conditional distribution.

To identify the appropriate conditional distribution, we note that a
reduced effective data model may at this point be written in terms of
the following signal-plus-noise residual,
\begin{equation}
  r = \M_{\mathrm{stars}}\a_{\mathrm{stars}} + \n_{\mathrm{w}}. 
\end{equation}
Since $\n_{\mathrm{w}}$ denotes white Gaussian noise, and $\M$ is a fixed matrix, this is a perfectly linear data model in $\a$, and the appropriate conditional probability distributions, $P(\a_{\mathrm{bright}}|\dv,\ldots)$ and $P(\a_{\mathrm{diff}}|\dv,\ldots)$, are both Gaussian. The appropriate sampling equation is therefore \citep[see, e.g., Appendix~1 in][]{bp01}
\begin{equation}
  \a_{i} = (\M_i^T \N^{-1} \M_i)^{-1} \M_i  \N^{-1}(\r-\M_j\a_j),
  \label{eq:amp_sampler}
\end{equation}
where $\{i,j\}$ now denote respectively bright or diffuse sources in a given Gibbs step, and then the reverse combination in the next Gibbs step. This algorithm is then iterated until convergence, and interspersed with similar Gibbs steps for all other free parameters in Eq.~\ref{eq:model}.

\subsection{Bright stars}
\label{sec:starmodel}

To complete the above algorithm, we need to specify $\M$, $\t$, and $\n_{\mathrm{s}}$, and we start by defining the catalog of bright stars. The first step in this procedure is to threshold the AllWise point source catalog \citep{wiseCat} at magnitude 7 in the W1 band. Many of these are located spatially very close to each other, and given the wide DIRBE beam this results in strong degeneracies in the fitted parameters of nearby objects. Furthermore, as described above, the computational engine of the current analysis framework is a Gibbs sampler, which is known to have a long correlation length for strongly correlated parameters \citep{geman:1984}. For this reason, we divide the set of bright objects into three separate sampling groups, within each of which no two objects are closer than a given distance. Specifically, each set is constructed by including the brightest remaining object, but only if it is further away from any already included source than 5'. All amplitudes within each group are sampled jointly with sparse matrix operators for the coupling matrix, $(\M_i^T \N^{-1} \M_i)^{-1}$, in Eq.~\ref{eq:amp_sampler}, resulting in a vanishing correlation length internally in each group.

For each star in these sampling groups we extract estimates of the effective temperature, $T_{\mathrm{eff}}$, the gravitational acceleration, $\log g$, and the metallicity, [M/H], from the \GAIA\ DR2 database \citep{gaiaCat}, and use those to identify a best-fit spectral energy density (SED) from the PHOENIX starlight database \citep{Husser_2013}. The distributions of these parameters for the selected stars are shown in Fig.~\ref{fig:gaiacat}. Unfortunately, the PHOENIX catalog lacks information for $T_{\mathrm{eff}} > 12000$\,K, and so for these we use the closest match; this accounts for 1.1\,\% of all bright WISE objects. In principle, this may lead to the SEDs of the hottest stars being biased low, but this effect is at least partially mitigated by the fact that we fit the overall amplitude of each star freely in the following analysis. 

\begin{figure}
\includegraphics[width=0.85\columnwidth]{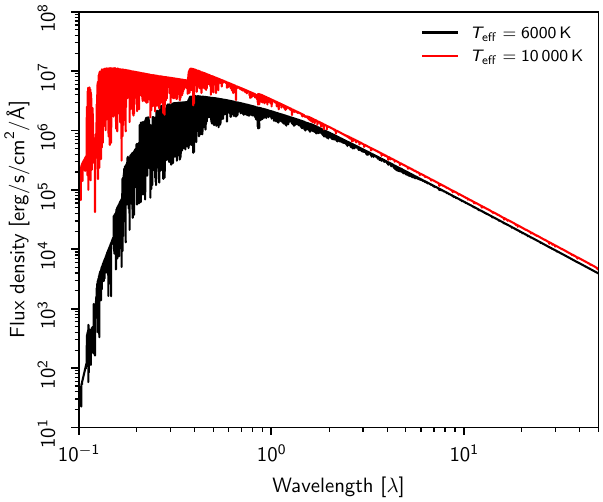}
\includegraphics[width=0.85\columnwidth]{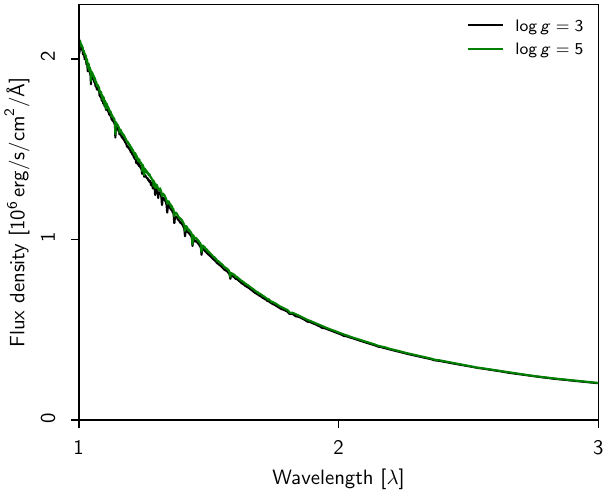}
\includegraphics[width=0.85\columnwidth]{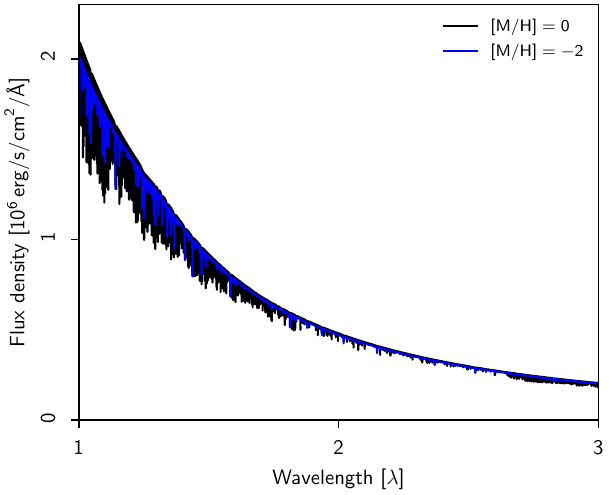}
  \caption{Comparison of PHOENIX spectra for different parameter compbinations. In each panel the black curve shows a reference star spectrum with $T_{\mathrm{eff}}= 6000$K, $\log g = 3.0$ and $[M/H]= 0.0$. The colored curve shows, from top to bottom, the resulting spectra when setting $T_\mathrm{eff}=10\,000\,\mathrm{K}$, $\log g = 5.0$, and $[M/H]= -2.0$. The spectra shown in the middle panel have been smoothed to highlight the broad features that are mostly relevant for the current analysis.}
  \label{fig:catalogueSEDs}
\end{figure}

\begin{figure}
\includegraphics[width=\columnwidth]{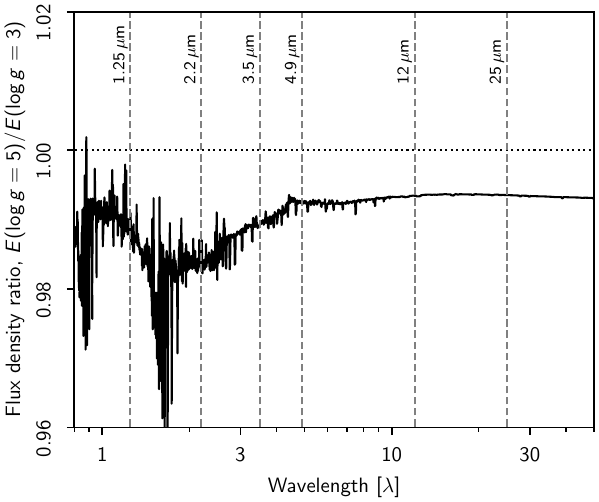}
\caption{Ratio between the PHOENIX spectra shown in the middle panel of Fig.~\ref{fig:catalogueSEDs}, corresponding to $E(\log g = 5)/E(\log g  = 3)$. The vertical dashed lines indicate the positions of the six shortest wavelength DIRBE channels.}
  \label{fig:logg_ratio}
\end{figure}

The relationship between the physical stellar parameters and the resulting star SEDs is illustrated in Fig.~\ref{fig:catalogueSEDs}. The black curves in each panel shows the SED for a typical star with $T_{\mathrm{eff}}= 6000$\,K, $\log g = 3$ and [M/H]\;$= 0$, while the colored curves show the SEDs resulting by changing one parameter at a time. As expected, the effective temperature has the biggest effect on the overall SED, especially when averaged over wide observing bands like those of DIRBE, while the other two parameters only have a relatively mild impact. This is further illustrated in Fig.~\ref{fig:logg_ratio}, which shows the ratio between two spectra with $\log g = 5$ and 3, and with the DIRBE channel center frequencies indicated as vertical dashed lines. Here we see that neglecting $\log g$ would lead to a 1-2\,\% relative bias between the 1.25 and 3.5\,$\mu$m channels.

\subsubsection{The PHOENIX Database}
\label{sec:phoenix}

To build an optimal SED model for each stellar source given the parameters retrieved from \GAIA, a few post-processing steps are needed for the spectra provided by the PHOENIX database. First, PHOENIX provides spectra in finite resolution bins, as summarized in Table~\ref{tab:phoenix}, whereas the parameters estimated from the \GAIA\ data have arbitrary precision. Instead of simply selecting the closest spectrum to each star, we compute a linear combination of the nearest six spectra, weighted by distance from the measured value, as given by 
\begin{align}
s &= \frac{1}{3}\bigg[\frac{s_{T-} (T_{+} - T) + s_{T+} (T - T_{-})}{T_{+} - T_{-}} \nonumber \\ 
  &\quad+ \frac{s_{g-} (g_{+} - g) + s_{g+} (g - g_{-})}{g_{+} - g_{-}} \nonumber \\
  &\quad+ \frac{s_{M-} (M_{+} - M) + s_{M+} (M - M_{-})}{M_{+} - M_{-}}\bigg].
\label{eq:starspec}
\end{align}
Here, $s$ represents a single spectrum, and we have used $T$, $g$ and $M$ as short forms for $T_{\mathrm{eff}}$, $\log g$ and [M/H]. The subscripts + and $-$ correspond to the closest reference value above and below the measured value (for example, 0.0 and 0.5 if $\log g=0.315$). 

\begin{figure}
  \centering
  \includegraphics[width=\columnwidth]{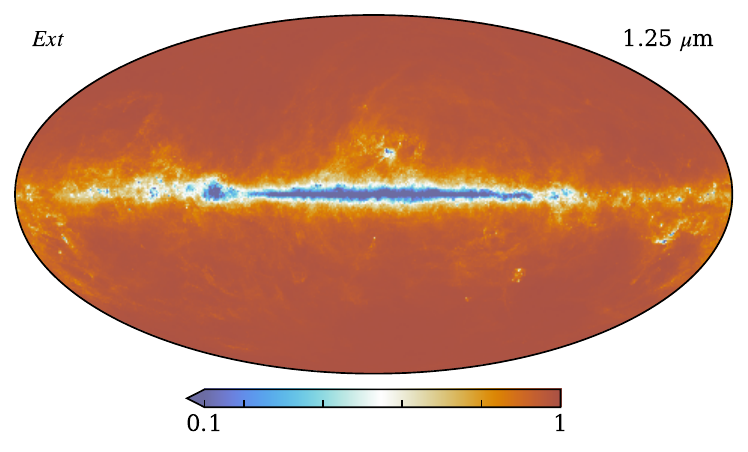}\\
  \caption{The full-sky extinction template, $\epsilon$, used in this analysis, evaluated for the DIRBE $1.25\mu$m band.}
  \label{fig:ext_template}
\end{figure}

\begin{table}
    \centering
    \newcolumntype{C}{ @{}>{${}}r<{{}$}@{} }
    \begin{tabular}{l c c c c c }
    \hline
    \hline
     Param & Step Size & \multicolumn{2}{c}{Old} & \multicolumn{2}{c}{Updated}\\ 
     & & Min & Max & Min & Max\\
    \hline
    \hline
    $T_{\mathrm{eff}}$ (K) & 50-100 & 2300 & 12000 & 3000 & 11900\\
    $\log g$ & 0.5 & 0.0 & 6.0 & 3.0 & 5.0 \\
    $[M/H]$ & 0.5 & -4.0 & 1.0 & -2.0 & 1.0 \\
     \hline
    \end{tabular}
    \caption{Parameters of spectra provided by the PHONEIX starlight database, both for the original and updated spectra.}
    \label{tab:phoenix}
\end{table}

A second complicating factor is that the updated release of the catalogue\footnote{\href{https://www.astro.uni-jena.de/Users/theory/for2285-phoenix/grid.php}{\texttt{https://www.astro.uni-jena.de/Users/theory/\newline for2285-phoenix/grid.php}}} contains only a subset of the files in the original catalogue, although they are believed to be more accurate. For stars in the extended parameter range, most importantly those with $\log g<3.0$, we therefore calculate spectra using the older files. 

Unfortunately, the older files are only defined up to $\lambda=5.5\,\mu$m, compared to the newer files, which extend to $\lambda=50\,\mu$m. Amplitude estimates for these stars in the $12\,\mu$m and $25\,\mu$m bands must therefore be determined by first, computing the spectra with the old PHOENIX files and the new PHOENIX files, using the closest endpoint ($\log g=3.0$ in the most common case). Then, we compute the ratio between the amplitudes of the $3.5\,\mu$m and $4.9\,\mu$m channels, and exploit the observation in Fig.~\ref{fig:logg_ratio} that the spectrum ratio is nearly constant above 4.9\,$\mu$m to extrapolate to longer wavelengths.

\subsubsection{Binary Stars}

About 25\% of the stars brighter than magnitude 12 were found to be binary stars in the \Gaia\ database, and for many of these, physical parameters are provided for each star separately. When modelling these in DIRBE, which has a low angular resolution, the two contributions must be averaged into one effective SED. For these, we first calculate the spectrum of each component according to Eq.~\ref{eq:starspec}. Since the flux values are given per unit of area, we then weight the two spectra according to an estimate of the area of the corresponding star, so that the total flux estimated from the binary system reads
\begin{equation}
    s_{\mathrm{b}} = \frac{s_1 r_1^2+ s_2 r_2^2}{r_1^2 + r_2^2},
\end{equation}
where $s_{1, 2}$ and $r_{1, 2}$ denote the flux and radius of the respective star in the binary system. The radii are not given explicitly in the \GAIA\ catalogue, so we therefore estimate their radius as \citep{kuiper_1938}
\begin{equation}
r = \exp\left(\frac{3.5\cdot 4}{5(3.5 \log(T_{\mathrm{eff}}) - \log(g))}\right ).
\end{equation}

\subsubsection{Extinction}
\label{sec:extinction_model}

At the highest frequencies, dust extinction becomes important near the Galactic plane. In this paper, we model this effect by $\epsilon(\nu)$, which we define as the ratio between the observed and the intrinsic flux from a given star, i.e., comparing the flux with and without dust extinction. We construct a simplified model for this based on the work by \cite{ext_model} and \cite{fitzpatric_19}. Specifically, we implement their 47-parameter model that extends from UV to infrared frequencies\footnote{Note that we believe there is a typo in \cite{ext_model}, which has added an additional negative sign before their value of $b_{IR}\ \alpha_{1}$. The correct value should be 1.06099.}, which provides a general sky-averaged estimate of $\epsilon(\nu)$. 

To account for spatial varations, we combine this averaged spectral behaviour with the $E(B-V)$ map from the \Planck\ 2013 dataset \citep{planck_extinc_2013}, starting from Eq.~1 of \cite{ext_model},
\begin{equation}
\frac{A(\lambda)}{A(V)} = a(\lambda) + b(\lambda)\left[ \frac{1}{R(V)} - \frac{1}{3.1}\right].
\end{equation}
Here, $\lambda$ is wavelength, $A$ is the absolute extinction, and $a$ and $b$ are the model coefficients listed by \citet{fitzpatric_19}. Finally, $R(V)$ is given by
\begin{equation}
R(V) = \frac{A(V)}{E(B-V)},
\end{equation}
and therefore denotes the ratio of absolute to selective extinction in the V-band. Substituting this relationship into the above equation, and using the full sky average case of $R(V) =3.1$ \citep{fullsky_ext}, allows us to omit the $b$ terms, and we simply obtain
\begin{equation}
A(\lambda) = 3.1\, E(B-V) a(\lambda).
\end{equation}
Finally, we convert from magnitudes to the linear $\epsilon(\nu)$ scaling factor to obtain
\begin{equation}
\epsilon(\lambda) \equiv \frac{F_{\mathrm{ext}}}{F_{\mathrm{raw}}} = e^{-\frac{3.1}{2.5} E(B-V) a(\lambda)}.
\end{equation}

The resulting spatial extinction template, evaluated for the 1.25$\mu$m band, can be seen in Fig.~\ref{fig:ext_template}. The majority of the extinction is occurring in the Galactic plane, but additional contributions can be seen tracing dust structures at high galactic latitudes. The template scales according to the relations shown in Fig.~7 of \cite{ext_model}, and is applied to the stars component as described in Eq.~\ref{eq:datamodel}.

\begin{figure}
  \centering
  \includegraphics[width=\columnwidth]{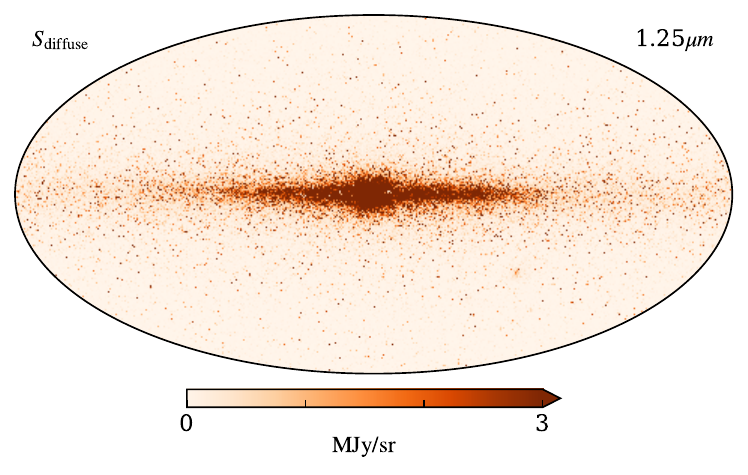}\\
  \caption{The unsmoothed template of diffuse star emission evaluated in the DIRBE 1.25$\mu$m band. This template is the same for all channels, but with a different overall amplitude.}
  \label{fig:diffuse}
\end{figure}

\begin{figure}
  \centering
  \includegraphics[width=0.7\linewidth]{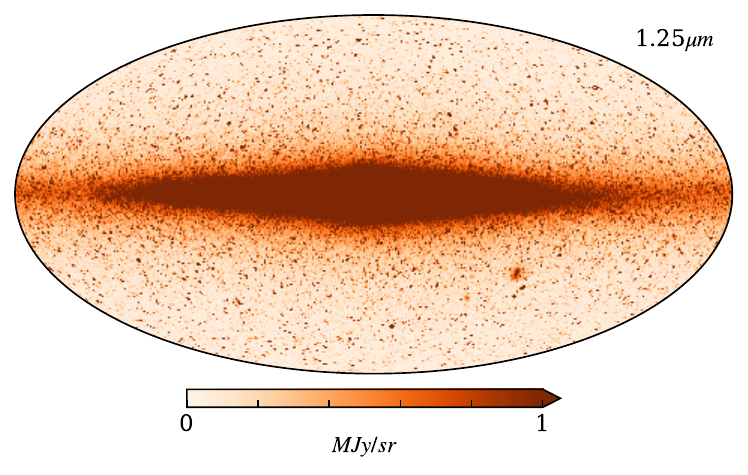} \\
  \includegraphics[width=0.7\linewidth]{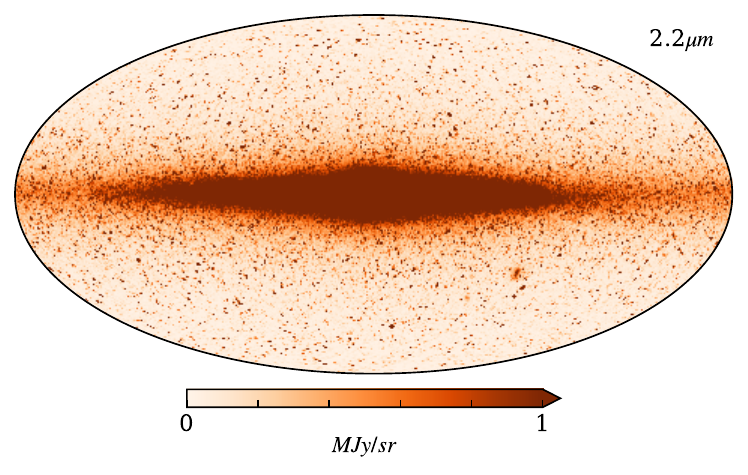} \\
  \includegraphics[width=0.7\linewidth]{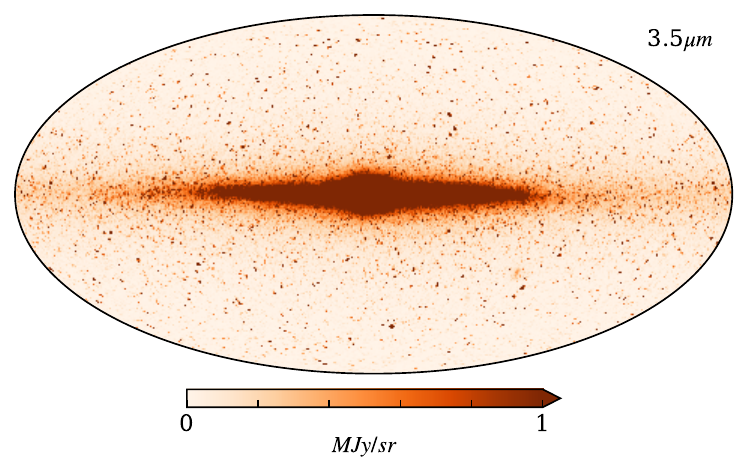} \\
  \includegraphics[width=0.7\linewidth]{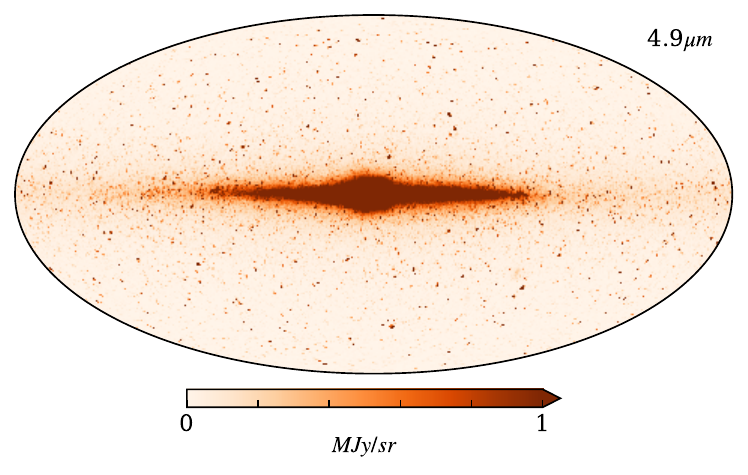} \\
  \includegraphics[width=0.7\linewidth]{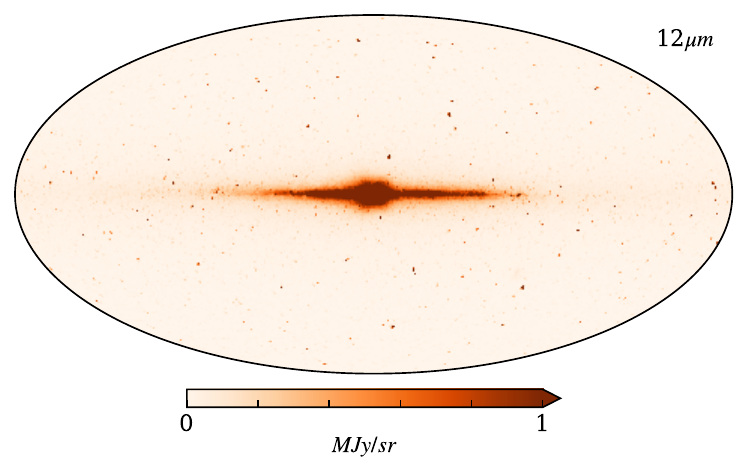} \\
  \includegraphics[width=0.7\linewidth]{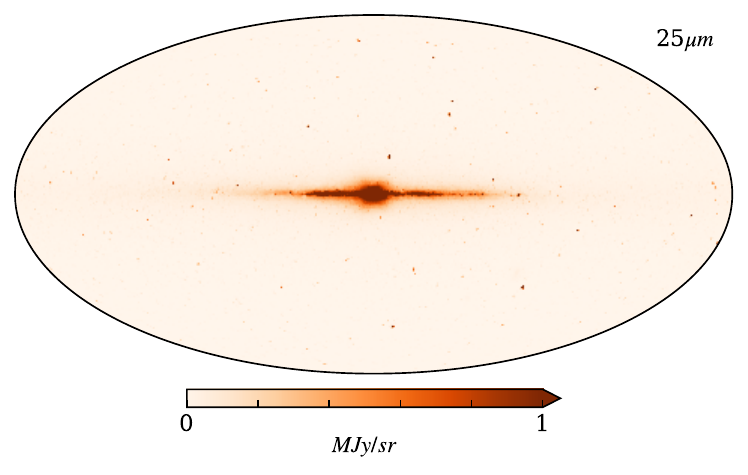} 
  \caption{Posterior mean star maps using the 1090 data samples from the \Cosmoglobe\ DR2 release. These maps are the sum of the bright and diffuse components for the first six DIRBE bands.}
  \label{fig:starsT}
\end{figure}

\begin{figure*}
  \includegraphics[width=0.95\textwidth]{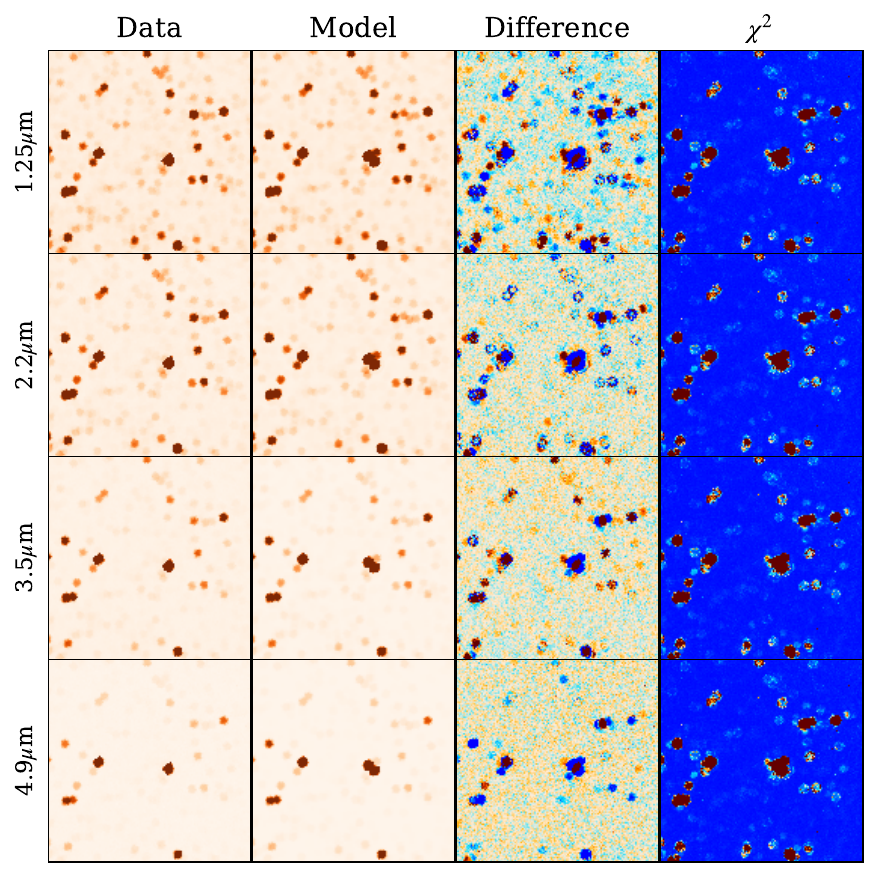}\\
  \vspace{-8pt}
  \hspace{1.2in}
  \includegraphics[scale=0.7]{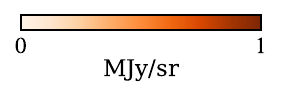}
  \hspace{0.95in}
  \includegraphics[scale=0.7]{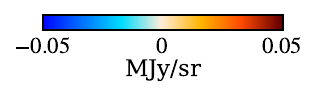}
  \hspace{0.15in}
  \includegraphics[scale=0.7]{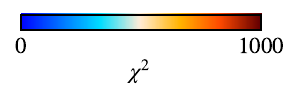}
  \caption{Zoom-ins on a typical subregion of the sky centred at $(20^\mathrm{o},70^\mathrm{o})$. The First and second columns show the map data and the total star model, respectively. The third column shows the (data-minus-model) difference between the two, and the final column shows the overall $\chi^2$ of the component separation fit.}
  \label{fig:zooms}
\end{figure*}

\begin{figure*}
  \centering
  \includegraphics[width=0.34\textwidth]{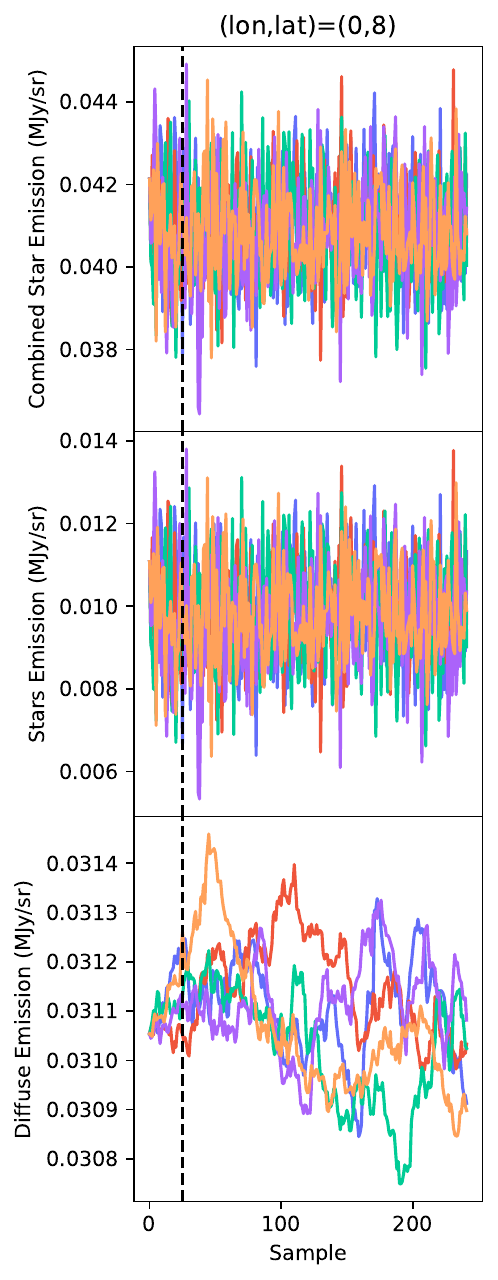}
  \includegraphics[width=0.31\textwidth]{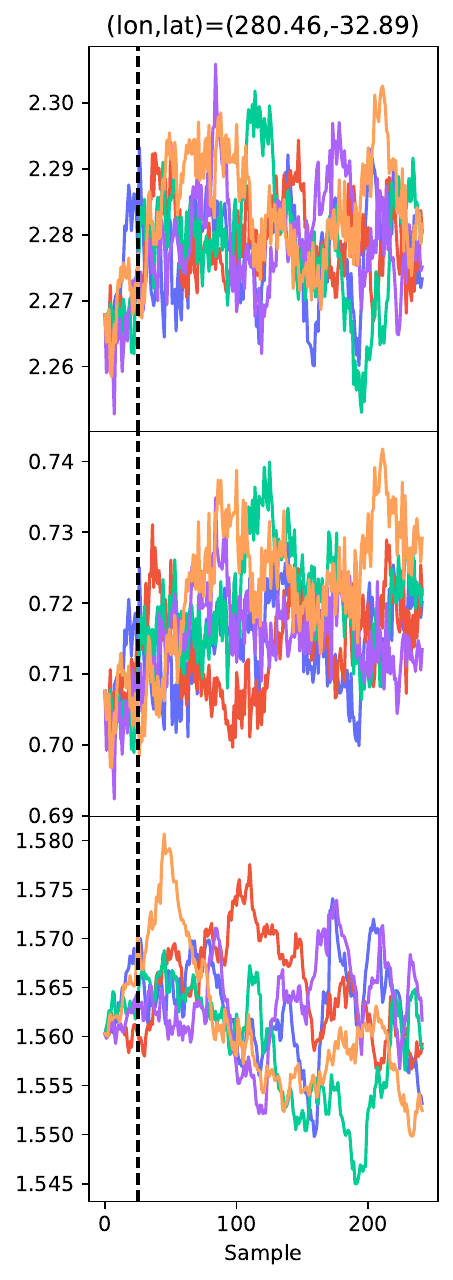}
  \includegraphics[width=0.31\textwidth]{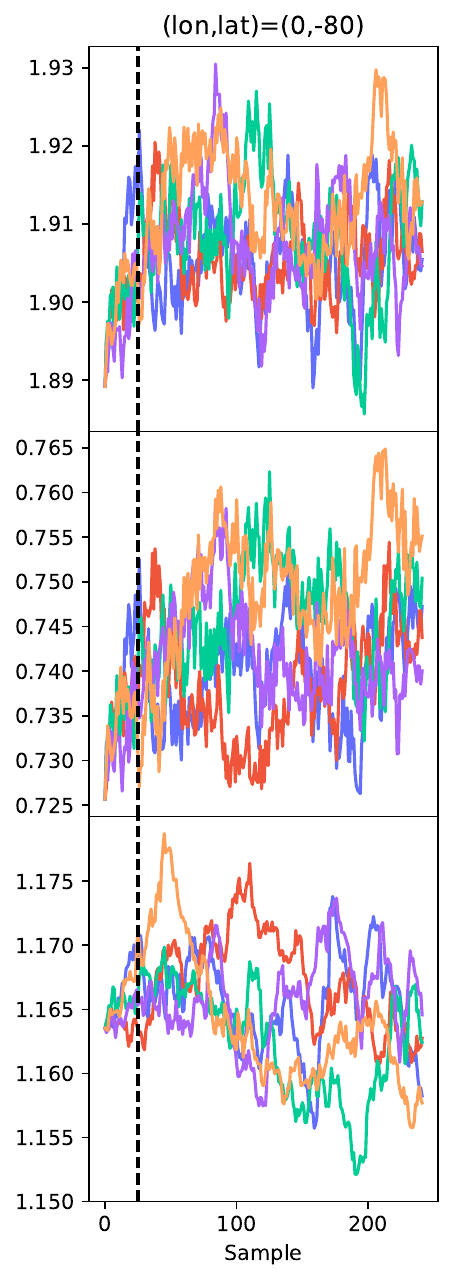}\\
  \caption{Trace plots of the total star amplitude in all five chains for three pixels in different regions of the sky (low latitudes, the Large Magellanic Cloud and high latitudes) in the DIRBE 1.25\,$\mu$m band. The dashed vertical lines indicate the end of the burn in period, and samples after that are included in our final sample set. The top row is the total star signal, and the bottom two rows are the bright stars and the diffuse template, respectively.}
  \label{fig:amptrace}
\end{figure*}

\begin{figure}
  \centering
  \includegraphics[width=\linewidth]{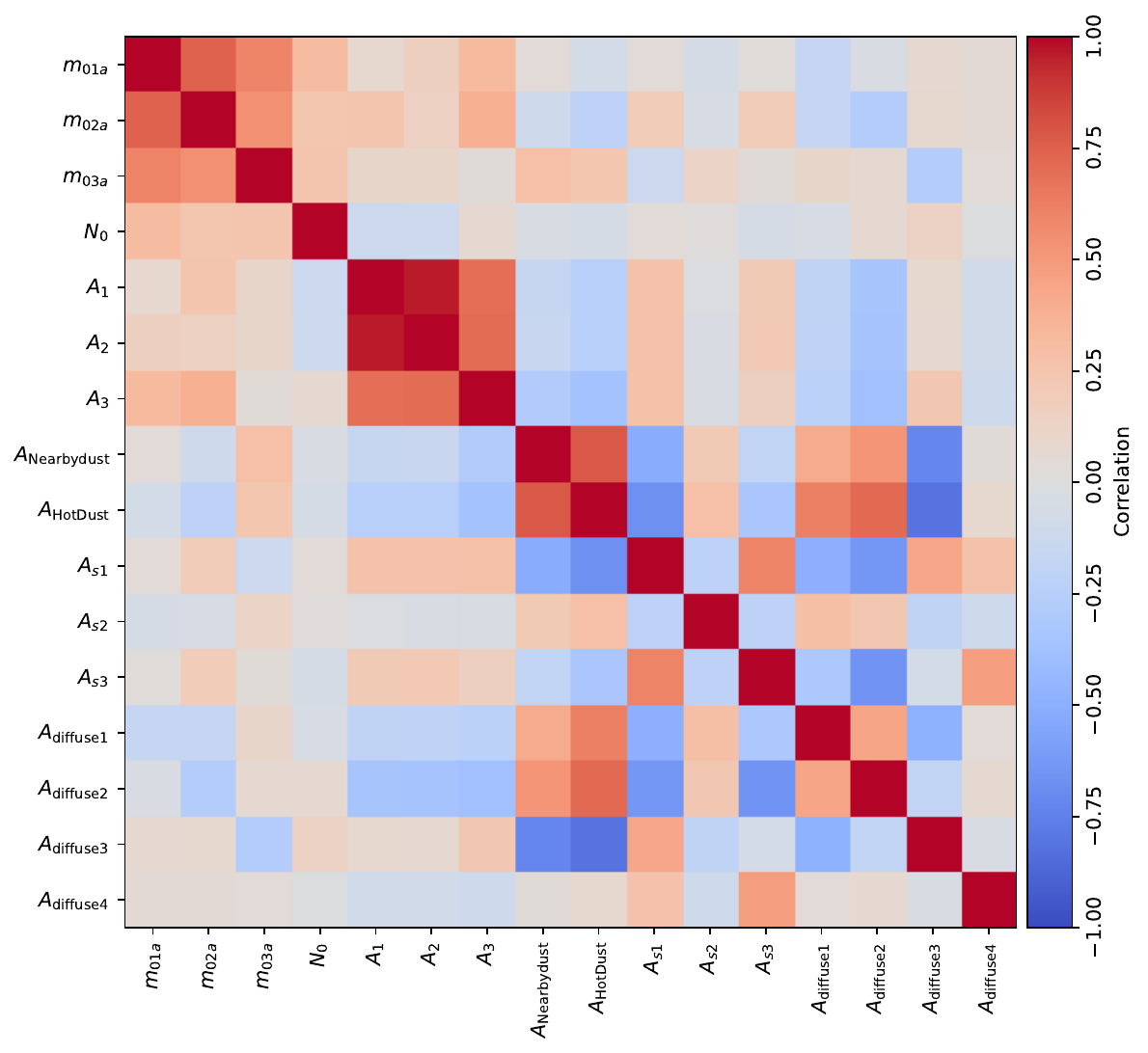}
  \caption{Correlation among a selection of parameters relevant for star modelling. From top to bottom, the included parameters are 1-3) the monopoles for DIRBE channels between 1.25 and 4.9\,$\mu$m; 4-7) zodiacal light cloud density, $N_{0}$, and albedoes, $\A_i$, for the 1.25-3.5\,$\mu$m bands; 8-9) full-sky dust amplitudes for the nearby and hot dust components; 10-12) the amplitudes of three randomly selected stars; and 13-16) the diffuse template amplitude for the 1.25-4.9\,$\mu$m bands.}
  \label{fig:corr}
\end{figure}

We conclude this section by noting that this extinction model is indeed simplified, as it does not take into account the actual 3D position of each star, but rather assume the \Planck\ $E(B-V)$ map to be valid for any position along a given line of sight. Because most of the high-latitude dust signal observed by \Planck\ originates from dust grains closer than a few hundred parsecs from the Sun, while the Galactic plane is filled with dust all through the Milky Way, this approximation is clearly much better at high than low Galactic latitudes. Because of this approximation, we exclude the 1.25\,$\mu$m channel when estimating $\a_{\mathrm{bright}}$, for which extinction is far more important than at longer wavelengths.

\subsection{Diffuse Star Emission}
\label{sec:diffusemodel}

After modelling the brightest sources using the method described above, we are left with a diffuse background of stars that are too dim to be individually resolved, but that in aggregate form a significant contribution to the total star signal at these wavelengths and so must be accounted for. We model this diffuse background using a single full-sky template, generated from the WISE W1 data at 3.4\,$\mu$m. We take the full AllWISE point source catalog, and select all sources with a valid W1 magnitude that are not already included in the bright point source model. To account for pixelization effects, we create this template at high resolution (HEALpix $N_{\mathrm{side}}=2048$), co-adding over all sources and then downgrade to $N_{\mathrm{side}}=512$ to match the resolution of the observed DIRBE maps. The resulting map is shown in Fig.~\ref{fig:diffuse}. This template is then included in the full sky model in Eq.~\ref{eq:model}, with an amplitude fitted separately for each channel.

\section{Astrophysical results}
\label{sec:results}

The model described in the previous section is fitted jointly to the DIRBE data together with Galactic dust, zodical light, and other components as described by \citet{CG02_01} and \citep{CG02_07}. A model of the DIRBE instrument is used to estimate other data artifacts, the most prominent of which was a sun-stationary signal of unknown origins. Following our previous work within the \Cosmoglobe\ framework \citep{bp01, watts2023_dr1}, we implement this model within the \texttt{Commander3} codebase \citep{bp03}, which is our Bayesian end-to-end analysis pipeline. The full posterior of all our model parameters was then sampled using Gibbs sampling, exploring the full space around the maximum likelihood values of each parameter. We computed a total of 1090 unique samples spread over 5 independent chains after discarding our 25 samples of burn-in data. 

In this paper, we present the results of the star models described in Sect.~\ref{sec:models}, and we direct interested readers to the companion papers for further information regarding the full framework. Specifically, details on the overall sampling approach can be found in \citet{CG02_01}; a new zodiacal light model is presented by \citet{CG02_02}; a novel multi-component dust model is described by \citet{CG02_05}, \citet{CG02_06} and \citet{CG02_07}; and the final limits on the CIB monopole derived from this work are presented by \citet{CG02_03}.

\subsection{Maps}

In Fig.~\ref{fig:starsT} we show the mean star maps for the full pipeline run in all six DIRBE bands where they are modelled. Here, the maps show the sum of the bright stars component and the diffuse background, which accounts for the  majority of the point source emission over the full sky. As expected, the emission is concentrated in the Galactic plane, but there is also an important contribution from high latitude stars.

In Fig.~\ref{fig:zooms} we show zoom-ins of a typical 20$^\mathrm{o}$ patch of sky centered at Galactic coordinates, $(l,b) = (20^\circ, 70^\circ)$, plotting the DIRBE sky map in each band for the four highest frequency bands where the stars are most significant. In the second column, we show the total star model for the same region, then the resulting residual, and finally the overall $ \chi^2$ of the fit. The residuals contain a variety of mis-modeled point sources.

We can identify three types of point source residuals in this third column of Fig.~\ref{fig:zooms}, although there are a large number of sources with residuals that are subdominant to the background noise level and so can be considered to be well modelled. Some of the point sources clearly show the imprints of spectral complexity, where the residual signature is blue in some bands and red in others, which implies that our knowledge of the spectra is imperfect and a single overall amplitude fit will always be insufficient to model that source. The second class of residuals appear to be ring shapes, where the center of the point source is well modelled, but we oversubtract emission around the edges. This could be caused by our simplistic modelling of the DIRBE beams, which relies on the scanning strategy to symmeterize the otherwise square beam profile of the DIRBE instrument. The third type of residuals is more complex, and look to be superpositions of multiple sources in the same region. These residuals could possibly be reduced by pruning point sources from the model that are too close to one another, or potentially by improving our models of the brightest sources. No attempt to do so was made in this work, as these residuals were only a small fraction of the total sky, and can be handled by masking using the $\chi^2$ values as a function of pixel, which clearly shows these same regions to be poorly modelled.

In Fig.~\ref{fig:amptrace}, we show the total amplitude of the stars in three pixels in different regions in the sky (low Galactic latitudes, the Large Magellanic Cloud, and high Galactic latitudes) as a function of iteration, for all five of our  independent sampling chains. The first 25 samples in each chain are discarded as burn in, and the remaining samples form the samples included in our analysis. We observe excellent Monte Carlo mixing in the top row, while the sky diffuse template amplitude mixes more slowly than the individual bright stars.

Figure~\ref{fig:corr} shows the correlation matrix computed over all 1090 accepted samples for a selected set of parameters, including monopoles for the 1.25-3.5\,$\mu$m bands (the 4.9\,$\mu$m monopole was not sampled, for details see \citet{CG02_03}); four zodiacal light parameters (the cloud density, $N_0$, and the albedos for the 1.25-3.5\,$\mu$m bands); two dust component amplitudes ($\A_{\mathrm{NearbyDust}}$ and $\A_{\mathrm{HotDust}}$); the amplitudes of three randomly selected stars ($\A_{\mathrm{s}i}$); and the diffuse starlight amplitude for the 1.25-4.9\,$\mu$m bands. The strongest positive correlations are seen internally among the monopoles and among the zodiacal light parameters; these are further discussed by \citep{CG02_01, CG02_02}. However, only a low level of correlations are observed between between these and the star model parameters. We do not see much cross correlation between the star parameters and other parameters in the global chain, with the exception of some of the individual star parameters and the dust amplitudes, which seem to vary. This suggests overall that the star parameters are only mildly degenerate with other full sky parameters.

\subsection{Spectral Energy Densities}

\begin{figure}
  \includegraphics[width=\columnwidth]{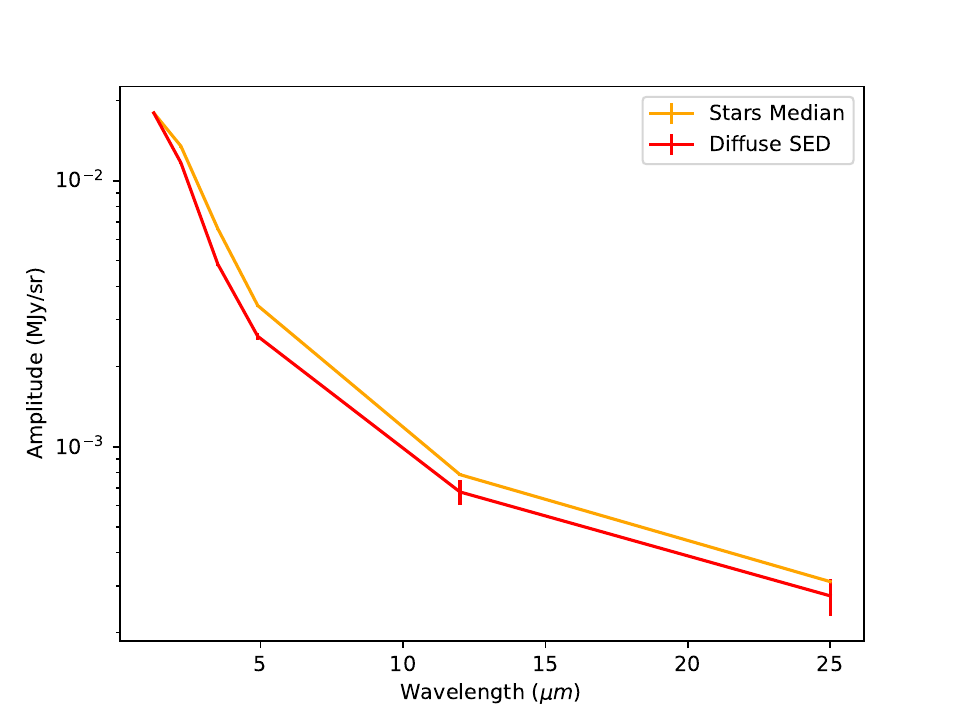}
  \caption{Star emission as a function of wavelength. The orange line shows the median individual star, averaged over the five chains. The variance on this median are too small to be seen on the plot. The red curve shows the mean of the diffuse template SED fit to the data, with the error bars showing the variance across the Markov chains.}
  \label{fig:starSEDs}
\end{figure}

In addition to the amplitude maps at each frequency, we can also examine the SEDs of the sources within our model. Figure~\ref{fig:starSEDs} shows the median SED model for all 424\ 829 bright star sources in our catalogue, as well as the SED for the diffuse template component, both normalized to a common amplitude in the 1.25\,$\mu$m band. These follow a similar overall shape, although the diffuse stars fall off faster than the bright stars with increasing wavelengths.

\begin{figure}
\includegraphics[width=\columnwidth]{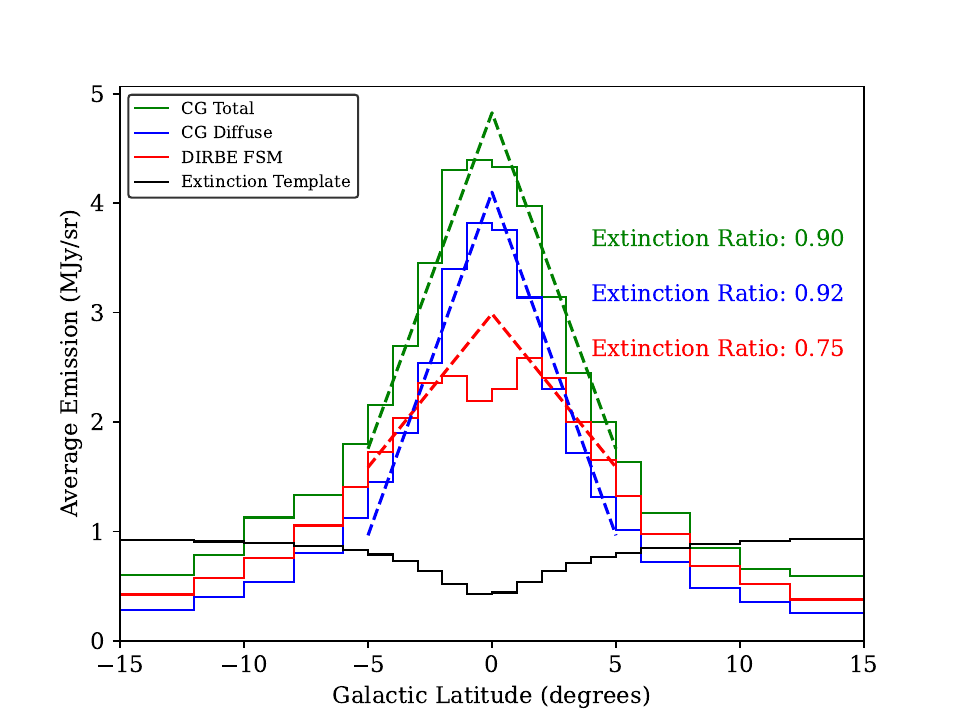}
  \caption{Profiles of the DIRBE Faint Star Model (FSM) and \Cosmoglobe\ star model as a function of galactic latitude. A linear extrapolation is overplotted in the core region from -5 to 5 degrees galactic latitude, but excluding the galactic center from -1 to 1 degree. The theoretical central value from this fit is compared to the measured values in the central bins to give estimates of the dust extinction in the galactic center for both cases.}
  \label{fig:extinction}
\end{figure}

\subsection{Extinction}

We also investigate our model of extinction, to determine how it compares to other estimates. Figure \ref{fig:extinction} shows the mean star emission as a function of Galactic latitude, zoomed in on the central region. The three curves shows the DIRBE Faint Source Model used in the legacy analysis \citep{dirbeFaint}, as well as our diffuse star model and the \Cosmoglobe\ total star emission estimate. Overplotted as dashed lines are linear fits to the data between 5 and 1 degrees, which, when extrapolated to the galactic center region, provides an approximation of the star signal in the absence of extinction. By comparing the peak amplitude of the fit to that of the models, we can approximate the magnitude of the extinction in the Galactic center, and this can be compared to the estimate from our extinction template (black line) which predicts about a 50\% extinction. We see that the FSM estimates a ratio of around 0.75 in the central region of the galaxy, while our model predicts about 0.9, which is likely an underestimate of the total extinction on the sky. In particular, because our extinction correction is only applied to the bright sources, the diffuse template contribution (which accounts for three quarters of the total power) at 1.25 $\mu m$ is too bright in the Galactic center, leading to an overall underestimate of the extinction. Future work will aim to improve on this limitation of our model by taking into account the full 3D position of each star.

 %previous estimates such as \cite{extinction} estimate it at 70\% at 1 $\mu$m and 94\% at 3.4 $\mu$m.

\section{Comparison to other Analyses}
\label{sec:consistency}

\begin{table*}
    \centering
    \newcolumntype{C}{ @{}>{${}}r<{{}$}@{} }
    \begin{tabular}{l c c c c c c c c c}
    \hline
    \hline
     Patch & $l$ & $b$ & $r$ & \multicolumn{2}{c}{Bright Stars} & \multicolumn{2}{c}{Diffuse Stars} & \multicolumn{2}{c}{Sum}\\ 
     & ($^{\circ}$) & ($^{\circ}$) & ($^{\circ}$) & \multicolumn{2}{c}{(kJy/sr)} & \multicolumn{2}{c}{(kJy/sr)} & \multicolumn{2}{c}{(kJy/sr)}\\
          &  & & & \cite{DIRBE2mass} & Current & \cite{DIRBE2mass} & Current  & \cite{DIRBE2mass} & Current\\
    \hline
    \hline
    \multicolumn{10}{c}{1.25$\mu$m}\\
    \hline
     1 \rule{0pt}{2ex} & 127.3 & 63.8 & 1.5 & 31.56 & 13.0 & 3.08 & 38.3 & 34.64 & 51.4\\
     2 & 107.7 & 57.7 & 2.0 & 42.72 & 18.8 & 3.62 & 44.5 & 45.34 & 63.3\\
     3 & 157.0 & -82.7 & 2.0 & 40.89 & 20.0 & 2.94 & 40.0 & 43.83 & 60.1\\
     4 & 257.8 & -59.4 & 1.9 & 51.88 & 28.6 & 3.92 & 45.3 & 55.8 & 73.9\\
     \hline
     \hline
     \multicolumn{10}{c}{2.2$\mu$m}\\
     \hline
     1 \rule{0pt}{2ex} & 127.3 & 63.8  & 1.5 & 20.79 & 9.0 & 1.58 & 25.7 & 22.37 & 34.7\\
     2 & 107.7 & 57.7 & 2.0 & 28.99 & 13.6 & 1.87 & 29.9 & 30.86 & 43.4\\
     3 & 157.0 & -82.7 & 2.0 & 28.62 & 15.1 & 1.50 & 26.9 & 30.12 & 42.0\\
     4 & 257.8 & -59.4 & 1.9 & 36.16 & 21.6 & 2.03 & 30.4 & 38.19 & 52.0\\
     \hline
    \end{tabular}
    \caption{Comparison of component intensities in selected fields from \cite{DIRBE2mass} and \Cosmoglobe\ at 1.25 and 2.2 $\mu$m. Each field is defined by Galactic longitude and latitude center coordinates, $(l,b)$, and a radius, $r$. The definitions of bright and diffuse sources differ between analyses, but the sum of the two allow reasonable comparison.}
    \label{tab:2mass}
\end{table*}

\subsection{DIRBE FSM}

\begin{figure}
\includegraphics[width=0.87\columnwidth]{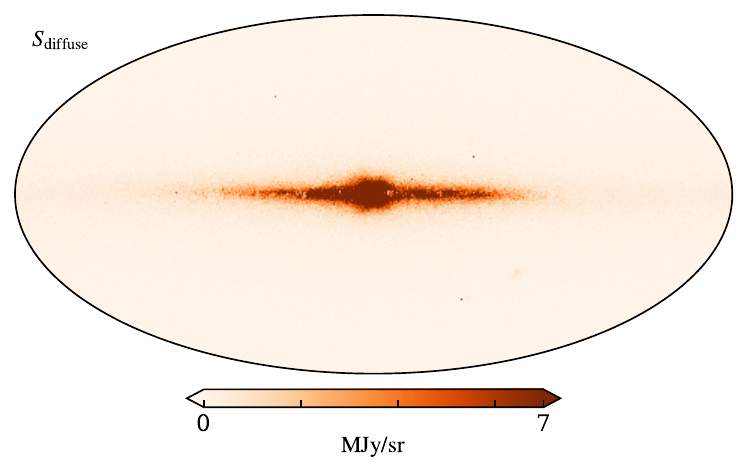}  \vspace{-4pt}\\
  \includegraphics[width=0.87\columnwidth]{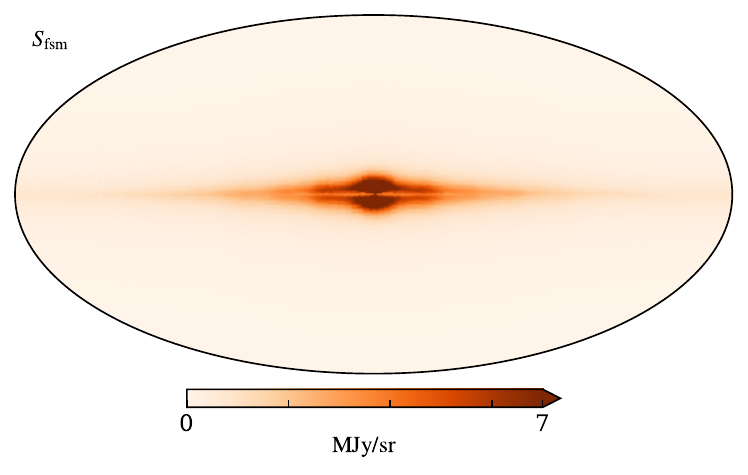}\\
  \includegraphics[width=0.87\columnwidth]{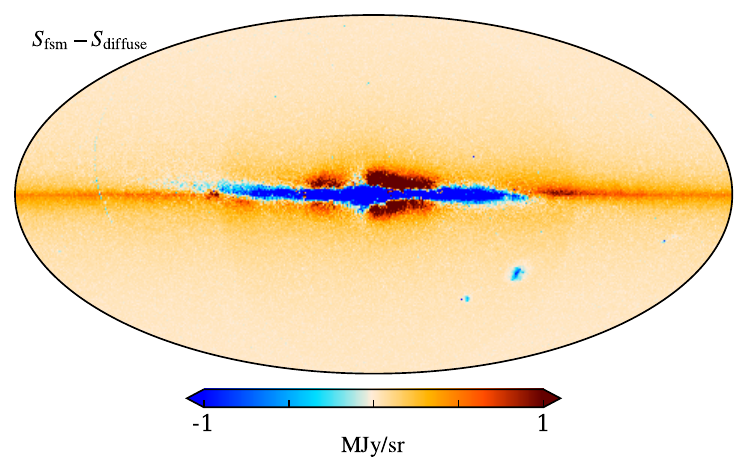}\\
  \includegraphics[width=0.87\columnwidth]{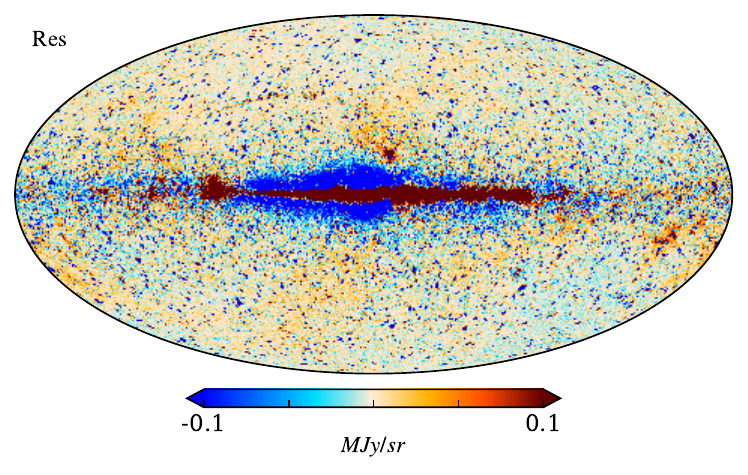}\\
  \caption{(Top panel:) \Cosmoglobe\ diffuse star template. (Second panel:) DIRBE FSM at 1.25\,$\mu$m retrieved from LAMBDA and converted from QuadCube to HEALPix. (Third panel:) Difference between FSM and our diffuse template. (Bottom panel:) \Cosmoglobe\ DR2 DIRBE 1.25\,$\mu$m residual map.}
  \label{fig:DIRBEfaint}
\end{figure}

The original DIRBE analysis team \citep{dirbeFaint} also removed point sources from their measurement of the CIB emission, but followed a different approach. Instead of modelling the bright sources, the original analysis simply masked pixels brighter than a threshold (15\,Jy at 1.25\,$\mu$m), which resulted in cutting almost 35\,\% of the sky at 1.25\,$\mu$m. For the diffuse sources, they build a model (the ``faint source model'' or FSM) using the methodology of \cite{wainscoat}, which integrates a model of source counts over the full sky, including models of Galactic morphology, 87 discrete source types, and the contribution from dust extinction. The top two panels of Fig.~\ref{fig:DIRBEfaint} compare our diffuse template with the FSM at 1.25\,$\mu$m, where the latter has been converted from a QuadCube pixelization to HEALpix, and plotted at $N_{\mathrm{side}}=256$. 

\begin{figure*}
  \centering
  \includegraphics[width=\textwidth]{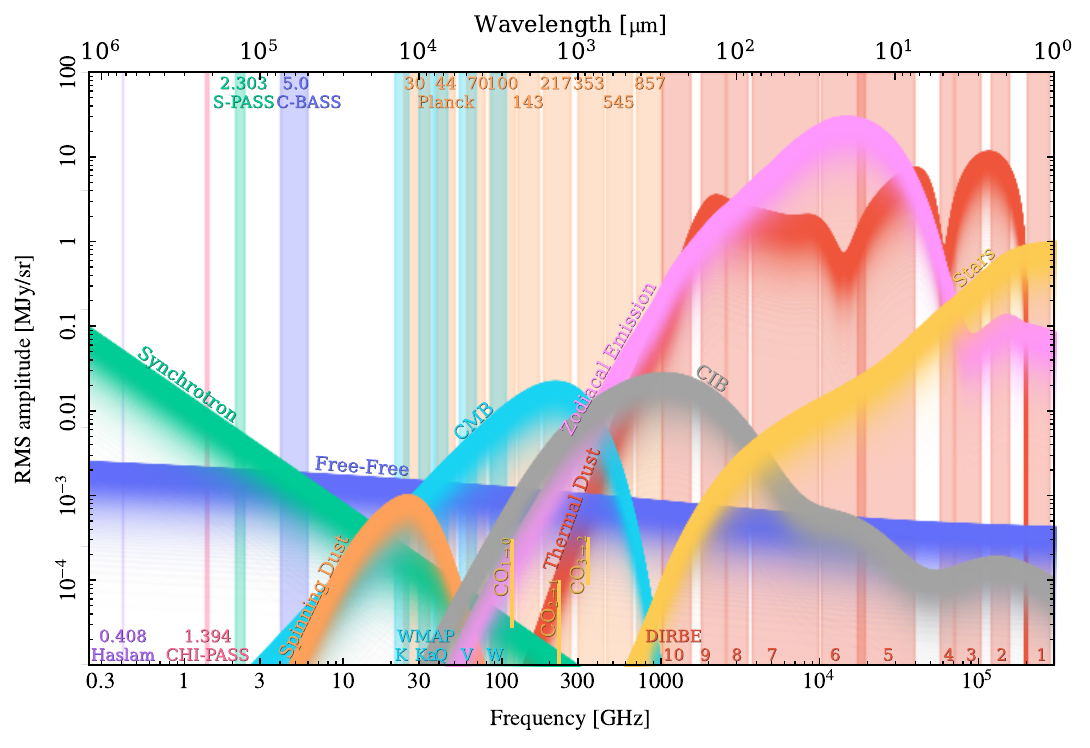}
  \caption{Components of the infrared temperature sky from 1$\mu$m to 1m wavelengths. The approximate contribution from stars is shown in yellow, and the observing bands from various experiments are indicated by the vertical lines. }
  \label{fig:sed}
\end{figure*}

The third panel shows the difference between the two templates. This map shows that our model, based on the integrated AllWISE point source catalogue, predicts more star emission than the DIRBE FSM in the Galactic plane where dust extinction is dominant. However, when we look at the 1.25\,$\mu$m residual map in the bottom panel, it is clear that the model is actually underpredicting the signal in this region, and more extinction is therefore disfavoured by the data.

In the Galactic bulge, our model predicts less emission than the FSM, which, when we look at the bottom panel of Fig.~\ref{fig:DIRBEfaint}, seems to be a better fit to the data. The blue residuals in the Galactic bulge show that the model overpredicts emission in general, and increasing the emission to the FSM level would result in a worse fit in this region. 

\subsection{DIRBE and 2MASS}

After the official DIRBE analysis, \cite{DIRBE2mass} (following \citealp{gorjian}) used the 2MASS catalogue to remove star emissions from DIRBE, aiming to derive an improved upper limit on the CIB monopole. Unfortunately, a positive detection was ultimately out of reach due to zodiacal light contamination. However, their analysis looked at four dark patches of the sky and used 2MASS to estimate the contribution of both bright (defined by a K-band magnitude lower than 14) and faint sources to the total signal in each patch. We compare these with the corresponding \Cosmoglobe\ estimates in Table \ref{tab:2mass}.

Differences in star classification between our analysis and \cite{DIRBE2mass} make it challenging to directly compare our amplitude estimates, but the sum of the bright and diffuse sources can still be compared. The \Cosmoglobe\ analysis predicts about 40\% more total star emission, depending on the field and the frequency, which is possibly due to having a deeper and more complete star catalogue than 2MASS. By successfully modelling and removing more star emission, we are able to improve upon their limits on the CIB, as discussed by \cite{CG02_03}. 

\section{Conclusions}
\label{sec:conclusions}

Stars are a critical component of the infrared sky, which must be appropriately modelled in order to avoid contaminating other components. This paper has presented a model of the star emission for the DIRBE data that is able to account for the vast majority of the star signal in these measurements. Using data from \Gaia\ and \WISE\, we constructed models of 424\,829 individual point sources as well as a diffuse background of the remaining 710\,825\,587 \WISE\ sources, and together these account for 91\% of the observed DIRBE flux density at 2.2$\mu$m, dropping to 1\% at 25$\mu$m.

Figure~\ref{fig:sed} shows the full sky average star SED (yellow band) derived from this work, plotted together with the other main sky components as a function of frequency and wavelength, from radio to infrared frequencies. The \Cosmoglobe\ DR2 sky model represents the first unified model of the microwave and infrared sky, and the star model presented here is an integral part of this work.

The star model showed robust Monte Carlo mixing properties within a Gibbs sampler, and it was able to fully describe the summed star emission in an effective manner. The resulting model was able to characterize and remove more star emission than earlier comparable works, in large part by using more modern data sets to trace star properties, and this consequently led to lower residuals and thereby stronger constraints on other astrophysical effects. Most notably, this work plays a critical role in a full-sky estimate of the CIB monopole that agrees with theoretical expectations around $1 \mu m$, and properly including the deep background of diffuse stars turned out to be the critical component in this effort. Indeed, in a earlier iteration of this work we obtained an estimate of the CIB monopole at 1.25\,$\mu$m that was in line with previous results, and significantly higher than expected from theoretical modelling, and after a careful re-assessment of the end-to-end pipeline, we discovered a bug in the diffuse template implementation which had omitted the dimmest objects. Properly accounting for those sources was precisely what led to the final strong limits on the CIB monopole presented by \citet{CG02_03}.

While already the current work represents a significant step forward, there are still many improvements that can and will be implemented in the near future, and perhaps the single most important effect regards dust extinction. Although this paper incorporates a first-order extinction correction, this is based on a full-sky template taken from the \Planck\ $E(B-V)$ template that is derived by comparison with distant quasar data. To get accurate numbers for stars that are embedded in the Galaxy, we need to incorporate 3D position information from \gaia\ to determine the line-of-sight extinction for each source. Constructing a 3D model of all the stars and gas in our Galaxy is beyond the scope of this paper, but is a top-priority issue for future work.

A second improvement concerns the beam treatment. Instead of assuming the DIRBE beam to be azimuthally symmetric, it would be much better to account for the unique square shape of the DIRBE response function through a time-domain algorithm such as deconvolution mapmaking \citep{artdeco}, and perform beam deconvolution (or symmetrization) per sample. This could potentially resolve some of the ``ring'' residuals around bright sources reported in this paper.

Next, more accurate SEDs should be established for stars with $T_{\mathrm{eff}}>12000$K. This may help reduce the residuals around the brightest sources, and this, combined with pruning nearby sources with vanishing amplitudes, could reduce the overall residual levels around the brightest sources. A more complete catalogue of star SEDs with higher resolution and expanded parameter ranges in the infrared would also improve the modelling of the individual sources, leading to lower residuals and fewer $\chi^2$ cuts.

Finally, including additional high frequency data directly in the full run would be an important next step for this work, and will be the subject of future \Cosmoglobe\ exploration. WISE is a prime candidate experiment for this work, but also AKARI \citep{akari}, \IRAS\ \citep{neugebauer:1984}, and SPHEREx \citep{Crill_2020} will provide key information. Incorporating these data directly into the existing \Cosmoglobe\ framework should provide enough extra information to constrain the star SEDs in the sampling stages instead of having to extrapolate them from other regimes. 

\begin{acknowledgements}
  MG would like to thank Prof.\ Alberto Dominguez for useful
  conversations that led to improvements in the star modelling and
  better limits on the CIB monopoles at 1.25 $\mu$m.  We thank
  Richard Arendt, Tony Banday, Johannes Eskilt, Dale Fixsen,
  Ken Ganga, Paul Goldsmith, Shuji Matsuura, Sven Wedemeyer, Janet
  Weiland and Edward Wright for useful suggestions and
  guidance.  The current work has received funding from the European
  Union’s Horizon 2020 research and innovation programme under grant
  agreement numbers 819478 (ERC; \textsc{Cosmoglobe}), 772253 (ERC;
  \textsc{bits2cosmology}), 101165647 (ERC, \textsc{Origins}),
    101141621 (ERC, \textsc{Commander}), and 101007633 (MSCA;
    \textsc{CMBInflate}).  This article reflects the views of the
    authors only. The funding body is not responsible for any use that
    may be made of the information contained therein. This research is
    also funded by the Research Council of Norway under grant
    agreements number 344934 (YRT; \textsc{CosmoglobeHD and 351037 FRIPRO; LiteBIRD-Norway}). Some of the
  results in this paper have been derived using healpy
  \citep{Zonca2019} and the HEALPix \citep{healpix} packages.  We
  acknowledge the use of the Legacy Archive for Microwave Background
  Data Analysis (LAMBDA), part of the High Energy Astrophysics Science
  Archive Center (HEASARC). HEASARC/LAMBDA is a service of the
  Astrophysics Science Division at the NASA Goddard Space Flight
  Center. This publication makes use of data products from the
  Wide-field Infrared Survey Explorer, which is a joint project of the
  University of California, Los Angeles, and the Jet Propulsion
  Laboratory/California Institute of Technology, funded by the
  National Aeronautics and Space Administration. This work has made
  use of data from the European Space Agency (ESA) mission {\it Gaia}
  (\url{https://www.cosmos.esa.int/gaia}), processed by the {\it Gaia}
  Data Processing and Analysis Consortium (DPAC,
  \url{https://www.cosmos.esa.int/web/gaia/dpac/consortium}). Funding
  for the DPAC has been provided by national institutions, in
  particular the institutions participating in the {\it Gaia}
  Multilateral Agreement.  We acknowledge the use of data provided by
  the Centre d'Analyse de Données Etendues (CADE), a service of
  IRAP-UPS/CNRS (http://cade.irap.omp.eu, \citealt{paradis:2012}).
  This paper and related research have been conducted during and with
  the support of the Italian national inter-university PhD programme
  in Space Science and Technology. Work on this article was produced
  while attending the PhD program in PhD in Space Science and
  Technology at the University of Trento, Cycle XXXIX, with the
  support of a scholarship financed by the Ministerial Decree no. 118
  of 2nd March 2023, based on the NRRP - funded by the European Union
  - NextGenerationEU - Mission 4 "Education and Research", Component 1
  "Enhancement of the offer of educational services: from nurseries to
  universities” - Investment 4.1 “Extension of the number of research
  doctorates and innovative doctorates for public administration and
  cultural heritage” - CUP E66E23000110001.
\end{acknowledgements}

%-------------------------------------------------------------
%                                       Table with references 
%-------------------------------------------------------------
%

\bibliographystyle{aa}
\bibliography{references, ../../common/CG_bibliography,../../common/Planck_bib}
\end{document}